\documentclass[preprint,12pt,authoryear]{elsarticle}

\usepackage{amssymb}
\usepackage{gensymb}
\usepackage{amsmath}
\usepackage{threeparttable}
\usepackage{booktabs}

\usepackage{lineno}

\journal{arXiv}

\begin{document}

\begin{frontmatter}



\title{Endogenous preference for non-market goods in carbon abatement decision}


\author[label1,label2]{Fangzhi Wang}
\author[label1,label2]{Hua Liao}
\author[label3,label4,label5,label6,label7,label8,label9]{Richard S.J. Tol}
\author[label10]{Changjing Ji}

\affiliation[label1]{organization={School of Management and Economics, Beijing Institute of Technology},Department and Organization
            city={Beijing},
            postcode={100081}, 
            country={China}}

\affiliation[label2]{organization={Center for Energy and Environmental Policy Research, Beijing Institute of Technology},Department and Organization
            city={Beijing},
            postcode={100081}, 
            country={China}}

\affiliation[label3]{organization={Department of Economics, University of Sussex}, 
            city={Falmer},
            postcode={BN1 9RH}, 
            country={UK}}

\affiliation[label4]{organization={Institute for Environmental Studies, Vrije Universiteit},
            city={Amsterdam},
            country={the Netherlands}}

\affiliation[label5]{organization={Department of Spatial Economics, Vrije Universiteit},
            city={Amsterdam},
            country={the Netherlands}}

\affiliation[label6]{organization={Tinbergen Institute}, 
            city={Amsterdam},
            country={the Netherlands}}

\affiliation[label7]{organization={CESifo},
            city={Munich},
            country={Germany}}

\affiliation[label8]{organization={Payne Institute for Public Policy, Colorado School of Mines},
            city={Golden},
            state={CO},
            country={USA}}

\affiliation[label9]{organization={College of Business, Abu Dhabi University},
            city={Abu Dhabi},
            country={UAE}}

\affiliation[label10]{organization={Institute of Carbon Neutrality, Shanghai Tech University},
            city={Shanghai},
            country={China}}

\begin{abstract}
Carbon abatement decisions are usually based on the implausible assumption of constant social preference. This paper focuses on a specific case of market and non-market goods, and investigates the optimal climate policy when social preference for them is also changed by climate policy in the DICE model. The relative price of non-market goods grows over time due to increases in both relative scarcity and appreciation of it. Therefore, climbing relative price brings upward the social cost of carbon denominated in terms of market goods. Because abatement decision affects the valuation of non-market goods in the utility function, unlike previous climate-economy models, we solve the model iteratively by taking the obtained abatement rates from the last run as inputs in the current run. The results in baseline calibration advocate a more stringent climate policy, where endogenous social preference to climate policy raises the social cost of carbon further by roughly 12\%-18\% this century. Moreover, neglecting changing social preference leads to an underestimate of non-market goods damages by 15\%. Our results support that climate policy is self-reinforced if it favors more expensive consumption type.
\end{abstract}




\end{frontmatter}


\section{Introduction}
\label{sec:sample1}

Climate change not only adversely impacts economic production, but also impairs non-market goods such as human health, clean water, biodiversity, etc \citep{Nordhaus1991EJ, Tol2009JEP,belval2023decision}. Optimal abatement decisions based on damage estimates should factor in both market goods and non-market goods. However, valuing non-market goods is inherently difficult, as it requires careful consideration of preferences, whose expression is endogenously determined in the complex climate-economy system. Furthermore, climate policy, like many other policies, can potentially influence social preferences \citep{bowles2012economic,mattauch2022economics}. This changes the value of non-market goods, and in turn affects the recommended climate policy. Thus, this paper investigates the optimal abatement decision in an integrated assessment model ("IAM" hereafter) when such policy alters social preferences for non-market goods.

By differentiating market from non-market goods, a strand of IAMs literature  \citep{tol1994damage,hasselmann1999intertemporal,hoel2007discounting,drupp2021relative,bastien2021use} consistently advocates a more stringent abatement policy. Market goods are growing overtime, whereas the absolute amount of some non-market goods are roughly stable (or even decreasing owing to, say, climate change) over the long run. The relative scarcity of non-market goods raises their price, and non-market goods are increasingly valuable. As climate change proceeds, increasing values of non-market goods can be associated with more expensive climate damages. Thus, it is recommended to adopt a stricter abatement policy based on cost-benefit analysis. Specifically, \citet{drupp2021relative} concludes that accounting for the rising price of non-market goods leads to a social cost of carbon about 50\% higher than Nordhaus suggested \citep{nordhaus2018projections}.
 
One assumption underneath this line of modelling is the \emph{fixed} social preference structure. By virtue of this, one can readily derive the relative price of non-market goods and assess the implications for optimal climate policy. However, as climate change is a long-run issue, social preferences need not be fixed. \citet{Beckage2022} and \citet{stern2022economics} criticized that IAMs fail to account for \emph{endogenous} preference. Hence, the carbon price recommended by these models lack credibility. \citet{peng2021climate} note that public opinion about climate policy is partly shaped by the success or failure of past climate policy, and that this feedback is currently lacking in IAMs. Public opinion is an expression of social preference. 

Endogenous preferences are generally assumed away in the broader climate management literature, two notable exceptions being \citet{konc2021social} and \citet{mattauch2022economics}. \citet{konc2021social} examined the level of carbon tax when consumers' preferences for products with different carbon intensities are influenced by peers in a social network. \citet{mattauch2022economics} investigated how to adjust carbon taxes in a static context when they crowd-in or -out social preference between dirty and clean consumption varieties. Both studies use examples of consumption with distinct carbon intensity. Their focus is not on valuing climate damages in climate policy. Their analysis is static. 

This paper endogenizes social preferences in an IAM. We consider the specific case where climate policy affects the social preferences for non-market goods. Policy design can increase public awareness and valuation of non-market goods \citep[e.g.:][]{christie2006valuing, kumar2012economics,tonin2019estimating}.

The paper takes two steps. First, it theoretically shows how changing preferences influence the relative price of non-market goods. A CES form utility function is utilized with time-varying weights for market and non-market goods. Weights attached to each variety of goods are assumed to reflect the associated social preferences. In addition to scarcity, the relative price of non-market goods will increase further with changing social preferences. Second, it establishes a dynamic climate-economy model, based on the seminal DICE model \citep{nordhaus2018projections} and its augmented version by \citet{drupp2021relative}, where social preferences are \emph{responsive} to abatement policy (unlike the original models). 

Notably, solving the proposed model is not as easy and direct as previous numerical decision exercises: Future preferences depend on current action while, as foresight is perfect, current action depends on future preferences. This decision process is lacking in previous climate-economy studies. This can be explained by two channels via which abatement policy affects social welfare. First, by reducing carbon emissions, global warming is curbed, and fewer climate damages materialize. Consequently, avoided damages serve to increase consumption levels of both market and non-market goods, but more so of the latter. Second, a hypothetical social planner is incentivized to attach a higher weight to the less expensive consumption type. However, this implies that a social planner can arbitrary decide social preference. The second channel is not only a moral issue, but also a matter of tractability in the real world. Thus, we need to rule out the second channel. To this end, we provide a novel solution method by iteration.

Numerical findings support a more stringent abatement policy when social preference for non-market goods is enhanced by this policy itself. First, with endogenous preferences, although the time of achieving net-zero emissions remains the same, unabated carbon emissions are greatly reduced before that (from 5.86 GtCO\textsubscript{2} to 1.29 GtCO\textsubscript{2} in 2050). Second, changing social preference for non-market goods increases the social cost of carbon from 124 to 139 US\$/tCO\textsubscript{2} (by 12\%) in 2020, and from 1981 to 2331 US\$/tCO\textsubscript{2} (by 18\%) in 2100. Third, by the end of this century, the model produces an optimal temperature rise of 2.5$^\circ C$, which is 0.80$^\circ C$ lower than DICE-2016. Sensitivity analyses are broadly consistent with the baseline results.

Our model can also be applied to value non-market goods damages, which is less common in previous IAM literature. We find that under optimal climate policy the climate impacts on non-market goods are equivalent to a consumption loss in market goods of 4 trillion US\$ in 2050 and 57 trillion US\$ in 2100, which account for 2.1\% and 11.5\% in aggregate market goods consumption, respectively. Moreover, because abstracting from endogenous preference fails to pin down the relative price of non-market goods correctly, it incurs an underestimate of non-market goods damages by 0.7 trillion US\$ in 2050 and 9 trillion US\$ in 2100.

The paper is related to two main strands of literature. First, it is closely tied to endogenous preferences in policy design, which is not recent \citep{bowles1998endogenous}, but is underexplored in climate economics and policy studies \citep{Beckage2022, stern2022economics, peng2021climate}. Second, it is a enhancement of the climate-economy model pioneered by \citet{nordhaus1992optimal}. To our knowledge, no paper explicitly considers endogenous preference in existing IAMs.

The remainder of the paper is structured as follows. Section \ref{sec:value} theoretically analyzes how time-varying preferences affect the valuation of non-market goods and hence social cost of carbon. Section \ref{sec:model} introduces the benchmark model, calibration process, and solution methods for our specific decision problems. Section \ref{sec:quan} presents numerical results. Section \ref{sec:conclusion} concludes the paper and discusses its findings.

\section{Non-market goods valuation under time-varying preference\label{sec:value}}
Our analyses below focus on two important concepts in climate economics and policy literature\textemdash relative price and social cost of carbon.

There is a representative household in the economy whose preference at time $t$ is defined as:
      \begin{equation}
      \label{eq:utility}
          U(C_t, E_t)= \left[\alpha_t E_t^{\frac{\theta-1}{\theta}} +(1-\alpha_t) C_t^{\frac{\theta-1}{\theta}}\right]^{\frac{\theta}{\theta-1}}
       \end{equation}
where $C_t$ and $E_t$ represent the amount of market and non-market consumption, respectively. $\theta$ is the substitution elasticity between two types of consumption. If $\theta \in (0,1)$, the two types of consumption are complementary; if $\theta \in (1,\infty)$, the two types are substitutes. For $\theta=1$, the utility function boils down to Cobb-Douglas form. $\alpha_t$ is the weight attached to non-market goods, and naturally $1-\alpha_t$ the weight for market goods. Following \citet{konc2021social} and \citet{mattauch2022economics}, one can interpret $\alpha_t$ as a taste parameter, the household's preference for non-market goods. The departure from previous studies is that the weight is here a time-varying, endogenous parameter, unlike a constant in \citet{drupp2021relative}\footnote{For simplicity, we abstract from the subsistence requirement noted by \citet{drupp2021relative}, which can be readily realized by adopting the Stone-Geary utility function.}. 

Thus, the implied willingness to pay for non-market goods is:
        \begin{equation}
        \label{eq:wtp}
          \frac{U_E}{U_C}=\frac{\alpha_t}{1-\alpha_t} \left(\frac{C_t}{E_t}\right)^{\frac{1}{\theta}}
       \end{equation} 
where $U_i$ denotes the first-order derivative of utility with respect to consumption $i$. The willingness to pay is alternatively called the relative price of non-market goods compared to market ones. As shown in Equation (\ref{eq:wtp}), in addition to the relative scarcity, the relative price is simultaneously influenced by the preference structure. To shed light on this, we turn to the relative price effect (abbreviated as "RPE" hereafter). According to \citet{hoel2007discounting} and \citet{drupp2021relative}, RPE reflects how the valuation of non-market goods changes with time. Given Equation (\ref{eq:wtp}), it reads as:
        \begin{align}
        \label{eq:RPE}
          RPE_t&=\frac{\frac{d}{dt}  \left(\frac{U_E}{U_C}\right)}{\frac{U_E}{U_C}}\notag\\
               &=\underbrace{g_{\alpha t} \times \frac{1}{1-\alpha_t}}_{preference \quad effect} + \underbrace{\frac{1}{\theta}(g_{Ct}-g_{Et})}_{scarcity \quad effect}
       \end{align}  
where $g_{\alpha t}$ is the change rate of the social preference for non-market goods. $g_{Ct}$
and $g_{Et}$ are growth rates of market and non-market goods, respectively. The first term on the right-hand side of Equation (\ref{eq:RPE}) is newly introduced in our study. It is positive as long as the social preference for non-market goods increases; and vice versa. Note that $\alpha_t$ always takes a value between 0 and 1, and hence $\frac{1}{1-\alpha_t}$ is positive. The second term is above zero if $g_{Ct}$ is strictly larger than $g_{Et}$. \citet{drupp2021relative} argued that because non-market goods remain generally unchanged, rapid accumulation of market goods renders an increasing relative price of non-market goods.

Our results add that the relative price effect is also adjusted by changing preferences. If the social preference for non-market goods is strengthened, $\alpha_t > \alpha_{t-1}$, the relative price will be increased further. On the contrary, if society is losing taste for non-market goods, the relative price effect due to scarcity comparison will be attenuated by the preference structure.

Increased relative price of non-market goods will accordingly drive up the social cost of carbon defined as:
\begin{equation}
    SCC_t=\frac{\frac{\partial U}{\partial M_t}}{\frac{\partial U}{\partial C_t}}
\end{equation}
where $M_t$ denotes carbon emission. The numerator reflects the impact of an additional tonne of carbon emission on utility that aggregates market and non-market damages. The denominator is the marginal value of market goods consumption. Thus, the climate impact on non-market goods is adjusted by relative price before being absorbed in the social cost of carbon. If the relative price of non-market goods is rising, the social cost of carbon will increase in tandem even though the physical damages remain unchanged.

\section{Model\label{sec:model}}
\begin{figure}
    \centering
    \includegraphics[width=1\textwidth]{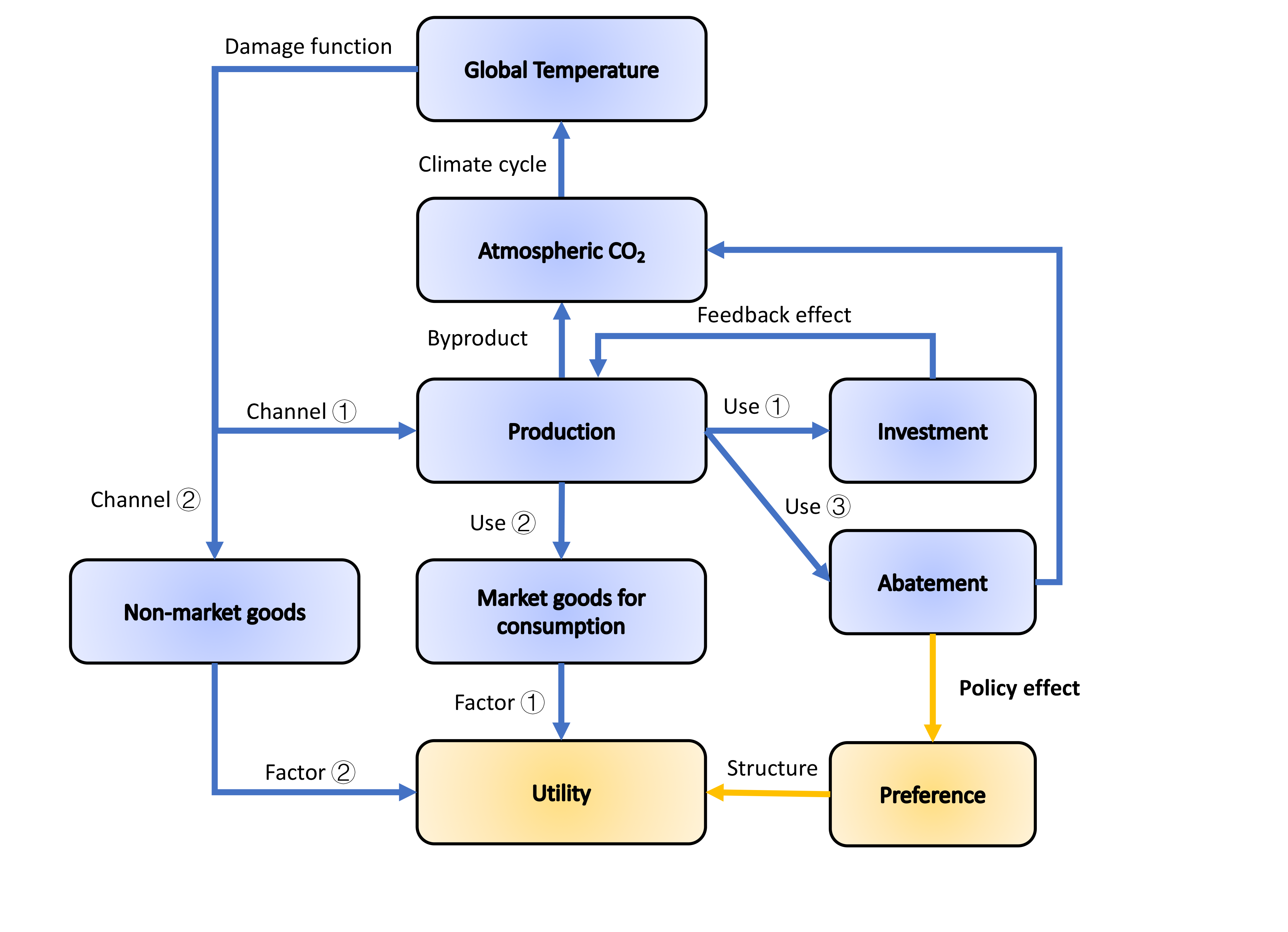}
    \caption{Framework}
    \label{fig:framework}
\end{figure}

The integrated assessment model we establish stands on the shoulder of the DICE model. Specifically, we build on the 2016 vintage \citep{nordhaus2018projections}\footnote{The most recent version is DICE-2023 \citep{NBERw31112}, but is still under revision at the moment. The core insight in this study would not change if we built on the latest version.}, introduce the relative price effect due to relative scarcity following \citet{drupp2021relative}, and then endogenize preferences. The framework of our model is presented in Figure \ref{fig:framework}. The yellow lines represent the new channel introduced in the model. In this section, we describe the main body of the model first, then introduce the calibration process, and conclude with the solution method.

\subsection{Main body}
The intertemporal welfare function is defined as:
        \begin{equation}
        \label{eq:lifetime utility}
          U \equiv \sum_{t=1}^{\infty} P_t \frac{1}{(1+\rho)^t} \frac{1}{1-\eta} \left[f(\alpha, \mu_t) E_t^{\frac{\theta-1}{\theta}} +(1-f(\alpha, \mu_t)) C_t^{\frac{\theta-1}{\theta}}\right]^{\frac{(1-\eta)\theta}{\theta-1}}
       \end{equation}
where $P_t$ is the exogenous population at period $t$, $C_t$ per capita consumption of market goods, $\rho$ the social time of preference, and $\eta$ the marginal utility of aggregate consumption.

$f(\alpha, \mu_t)$ is the weight attached to non-market goods, and is the counterpart of $\alpha_t$ in Equation (\ref{eq:utility}). Different from $\alpha_t$, it specifies that social preference is influenced by the abatement rate $\mu_t$, in addition to the fixed weight $\alpha$. Following \citet{mattauch2022economics}, we assume that the abatement rate exerts its impact in a linear form\footnote{\citet{bowles2012economic} studies alternatives to linearity. In addition, this paper is more interested in the long-term gradual changes in social preference and hence abstracts from climate-related decisions under uncertainty as \citet{webster2008incorporating}. All this is deferred to future research. } so that:
        \begin{equation}
        \label{eq:preference}
            f(\alpha, \mu_t)=\alpha+\beta_\mu*\mu_t
        \end{equation}
where $\beta_\mu$ measures the extent to which the abatement rate influences the preference for non-market goods. Previous studies \citep{christie2006valuing, kumar2012economics,tonin2019estimating} provided some evidence that proper policy designs lead to increased social appreciation of non-market goods. Motivated by these, we set $\beta_\mu>0$. This can be the case, for example, when part of revenues from abatement efforts are used in environment-related education or propaganda and consequently raise public appreciation of non-market goods. Alternatively, people may rationalize the carbon taxes that they pay by convincing themselves that they care more about climate change than they used to.

Global output net of climate damages and abatement costs is governed by:
        \begin{equation}
        \label{eq:prod}
             Y_t=(1-\Lambda_t)A_t K_t^\gamma P_t^{1-\gamma}/(1+\Omega_t)
        \end{equation}
where $A_t$ denotes total productivity level, $K_t$ capital stock and $P_t$ population. $\gamma$ is the capital share. $\Lambda_t$ is the abatement cost function, and  $1/((1+\Omega_t))$ measures the relative level of climate damages, both of which are fraction of aggregate economic output. Specifically, $\Omega_t$ is given by a quadratic function:
        \begin{equation}
        \label{eq:damage}
             \Omega_t=\psi_1 T_{t}^2
        \end{equation}
where $T_t$ refers to atmospheric temperature change relative to preindustrial levels. 

In line with \citet{hoel2007discounting} and \citet{drupp2021relative}, non-market goods cannot be directly produced \emph{à la} market goods\footnote{Non-market goods can alternatively be modelled as natural capital , that is used in economic production \citep{bastien2021use, bastien2022climate} or feeds into utility \citep{HeinzowTol2003, Brooks2014}. We focus on preferences and model natural services as a flow. A second capital good would complicate the dynamics and would best be introduced together with a control variable, such as afforestation or nature conservation.}, but are susceptible to climate impact:
        \begin{equation}
        \label{eq:non-market}
             E_t=\frac{E_0}{1+\psi_2 T_{t}^2} 
        \end{equation}
Thus, non-market goods decrease in accordance with increasing temperature.\footnote{When temperature falls, Equation (\ref{eq:non-market}) implies that non-market goods recover to previous levels. However, temperatures do not fall in any of our experiments.} 

The abatement cost function as a fraction of aggregate production in Equation (\ref{eq:prod}) is specified as:
        \begin{equation}
        \label{eq:abate}
             \Lambda_t=\varphi_{1t}\mu_t^{\varphi_2} 
        \end{equation}
where $\varphi_{1t}$ is the unit cost parameter. $\varphi_2 > 1$ mirrors that a higher abatement rate $\mu_t$ is associated with non-decreasing marginal abatement costs \citep{gillingham2018cost}.

On the demand side, net global output can be used either for consumption of market goods or investment:
        \begin{equation}
        \label{eq:allocation}
             Y_t=C_t + I_t
        \end{equation}
where $C_t$ denotes aggregate market goods consumption such that $c_t=C_t/P_t$. Consumption increases the current-period utility of the representative household. With new investment $I_t$ added at each period, physical capital accumulates according to:
        \begin{equation}
        \label{eq:capital accumulation}
             K_{t+1}=(1-\delta)K_t + I_t
        \end{equation}
where $\delta$ is the depreciation rate of physical capital.

Given the abatement rate $\mu_t$, industrial carbon emissions into the atmosphere are:
        \begin{equation}
        \label{eq:emission}
             M_{t}=(1-\mu_t)\sigma_t Y_t
        \end{equation}
where $\sigma_t$ denotes the exogenously-given carbon intensity, \emph{i.e.}, carbon content per unit of gross economic output.

The DICE-2016 model\footnote{For more details on climate modules, see \citet{nordhaus2013dice}.} adopts equations of the carbon cycle including three reservoirs (carbon in the atmosphere $L_t^{At}$, the upper oceans and the biosphere $L_t^{Up}$, and the deep oceans $L_t^{Lo}$):
\begin{equation}
    \begin{pmatrix} L_t^{At}\\L_t^{Up}\\L_t^{Lo} \end{pmatrix}=
    \begin{pmatrix} \phi_{11} & \phi_{21} & 0\\\phi_{12} & \phi_{22} & \phi_{32} \\ 0 & \phi_{23} & \phi_{33} \end{pmatrix}
    \begin{pmatrix} L_{t-1}^{At}\\L_{t-1}^{Up}\\L_{t-1}^{Lo}\end{pmatrix}+
    \begin{pmatrix} M_t+M_t^{Land}\\0\\0 \end{pmatrix}
\end{equation}
where $\phi_{ij}$ depicts how carbon flows between reservoirs and $M_t^{Land}$ are exogenous carbon emissions from land use. Both industrial and land use carbon emissions enter the atmosphere first, and are diffused into the other reservoirs.

Radiative forcing $F_{t}$ is jointly determined by the current level of atmospheric carbon concentrations compared to 1750 and the exogenous forcing from other greenhouse gases $F_{t}^{Ex}$:
\begin{align}
F_{t} = \kappa \left[ \ln \left( L_t^{At}/ L_{1750}^{At}\right) / \ln(2) \right] + F_{t}^{Ex}
\end{align}
where $\kappa$ is calibrated to the forcing of CO\textsubscript{2} doubling. 

Finally, radiative forcing raises the atmospheric temperature $T_t$ and, indirectly because of heat exchange, the deep ocean temperature $T_t^{Lo}$: 
\begin{equation}
    \label{eq:tem}
    \begin{pmatrix} T_t\\T_t^{Lo} \end{pmatrix}=
    \begin{pmatrix} 1-\zeta_1\zeta_2-\zeta_1\zeta_3 & \zeta_1\zeta_3  \\ 1-\zeta_4 & \zeta_4 \end{pmatrix}
    \begin{pmatrix} T_{t-1}\\T_{t-1}^{Lo} \end{pmatrix}+
    \begin{pmatrix} \zeta_1\chi_t\\0 \end{pmatrix}
\end{equation}
where $\{\zeta_i\}_{i=1}^4$ are the heat exchange parameters between the atmosphere and the ocean. Owing to diffusive inertia of different reservoirs, there are lags in the climate system.

\subsection{Calibration}
\label{sec:cal}
All parameters, unless otherwise stated, are borrowed from \citet{nordhaus2018projections} and \citet{drupp2021relative}. Table \ref{tab:para} displays the key parameters in this study. The initial monetary value of non-market goods (76 trillion, 2005 US\$) is considered to be the same as that of market goods. In addition, the fixed share of non-market goods in utility function is 0.1, and the substitution elasticity between two kinds of consumption is 0.5. \citet{nordhaus2018projections} aggregated both market and non-market damages in the DICE-2016 vintage, which totals 2.12\% loss of GDP at a temperature rise of 3$^\circ C$. The share of non-market damage is a guess of 20\%. By comparison, to account for the relative price effect, \citet{drupp2021relative} followed \citet{sterner2008rp} and assumed that non-market damage is comparable to market damages (50\% for each). Namely, each damage in effect accounts for a GDP loss of 1.63\%, and aggregate damage totals 3.26\% at a temperature rise of 3$^\circ C$. Recently, \citet{NBERw31112} revised the damage estimate to 3.12\% of global output, based on the climate impact studies surveyed by \citet{piontek2021integrated} as well as additionally adding 1\% for the impact of tipping point events \citep{dietz2021economic}. This revision is generally consistent with \citet{drupp2021relative} from which we maintain $\psi_1=0.00181$ and $\psi_2=0.016$. The relative shares are close to the results by \citet{Tol2022}, who finds that 45.7\% of total impacts are non-market.

\begin{table}[htbp]
  \centering
  \caption{Some key parameters}
    \begin{tabular}{p{5em}p{22em}p{3.5em}}
    \toprule
    Parameter & Note  & Value  \\
    \midrule
    $\alpha$ & Initial weight of NMG in utility function & 0.1  \\
    $\theta$ & Substitution elasticity between MG and NMG & 0.5  \\
    $E_0$ & Initial monetary value of NMG (trillion 2005 USD) & 77.74 \\
    $\psi_1$ & Damage coefficient on MG & 0.0181   \\
    $\psi_2$ & Damage coefficient on NMG & 0.016   \\
    $\beta_\mu$ & Ability of policies to alter social preference & 0.02  \\
    \bottomrule
    \end{tabular}%
  \label{tab:para}%
    \begin{tablenotes}[flushleft]
    \footnotesize
    \item Notes: MG is short for market goods, and NMG for non-market goods.  
    \end{tablenotes}
\end{table}

The most difficult parameter is the impact of abatement policies on social preference. Unfortunately, there are scant empirical studies for calibrating this parameter. In the baseline model, $\beta_{\mu}$ takes a guess value of 0.02, implying that a policy to reduce emissions by 10\% fosters an increase in the share of non-market goods in total expenditures by 0.2\%. In addition, we consider different values for sensitivity tests, ranging from 0 to 4\%. In most cases, positive values are selected for $\beta_{\mu}$ to reflect the intuition that proper policy design raises social awareness of non-market goods. We also consider a specific case $\beta = -0.01$ in the sensitivity analysis to include an extreme case that carbon abatement policies, if implemented inappropriately, may incur people to  resent preserving non-market goods.

\citet{nordhaus2018projections} calibrated the economic module such that annual GDP per capital growth averages 2.1\% from 2015 to 2050, and 1.9\% from 2050 to 2100. For the climate module, the mean warming is 3.1 $^\circ C$ for an equilibrium CO\textsubscript{2} doubling and the transient climate sensitivity is 1.7 $^\circ C$.

\subsection{Decision problem and solution methods}
We solve an optimal control decision problem that maximizes Equation (\ref{eq:lifetime utility}) subject to constraints (\ref{eq:preference})-(\ref{eq:tem}). However, note that the abatement rate $\mu_t$ also appears in the utility function (\ref{eq:lifetime utility}). Our understanding is \textbf{not} that a hypothetical social planner, by changing abatement policy, intentionally alters social preference. To see this, suppose for illustrative convenience that two types of consumption are prefect substitutes so that the instantaneous utility function can be rewritten to $U(c_t, E_t)=(\alpha+\beta_\mu \mu_t)(E_t-c_t)+c_t$. Two types of consumption are calibrated with the same value in the initial period. Also note that because market goods grow faster than non-market ones, it is then expected that $E_t-c_t \le 0$. Because $\beta_\mu$ is positive by assumption, any increment in $\mu$ leads to declining utility directly. If $\beta_\mu$ is enough high, the hypothetical social planner will not be motivated to reduce carbon emission at all, because the direct utility loss from preference changes outweighs the utility gain due to avoided damages. 

Therefore, instead, we solve the optimization problem as if preferences had always been like the new ones. That is, we solve the decision problem using the following iteration method:

\textbf{Step 1} Run the optimal scenario in the DICE model that explicitly differentiates market and non-market goods, and save a series of abatement rates in each period;

\textbf{Step 2} Use the abatement rates obtained in Step 1 to calculate a series of new weights according to $f(\alpha, \mu_t)$ for both market and non-market goods, re-run the model, and save the new abatement rates;

\textbf{Step 3} Repeat Step 2 until the obtained abatement rates between two subsequent runs are almost identical. 
    
For each run, the abatement rates appearing in the utility function are predetermined, so the model searches the optimal allocation as in the fixed preferences situation. However, because the abatement rates obtained in each run are utilized as initial inputs in the next run, they reshape the social preference structure and are endogenous in the model. The computational method is based on presumption that the obtained abatement rate in each iteration will converge to its equilibrium level, which has been testified by numerical exercises in this study. For example, in our baseline calibration ($\beta_\mu =0.02$), the solutions stabilize after four iterations.

\section{Quantitative results\label{sec:quan}}
We first present the numerical results where parameters are calibrated to their reference values as in Section \ref{sec:cal}. As $\beta_\mu$ is a core parameter in our model but lacks empirical supports, we additionally perform a sensitivity test.

\subsection{Baseline results}
We compare our baseline results to DICE-2016, which does not differentiate non-market goods explicitly, and \citet{drupp2021relative}, which only considers relative price effect but abstracts from endogenous preferences.

\begin{figure}[htp]
\centering
\begin{threeparttable}
    \begin{minipage}[t]{0.49\textwidth}
        \centering
        \includegraphics[width=1\textwidth]{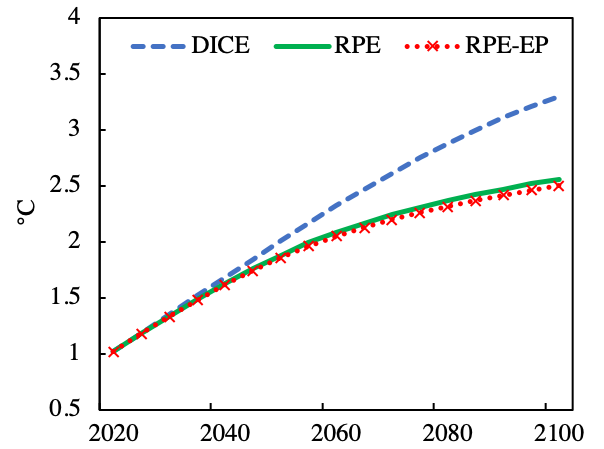}
        \centerline{\footnotesize{(a): Atmospheric temperature rise}}
    \end{minipage}
    \begin{minipage}[t]{0.49\textwidth}
        \centering
        \includegraphics[width=1\textwidth]{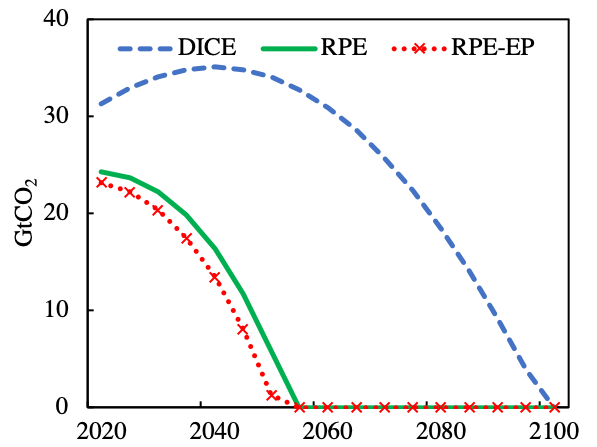}
        \centerline{\footnotesize{(b): Industrial emissions}}
    \end{minipage}
    \begin{minipage}[t]{0.49\textwidth}
        \centering
        \includegraphics[width=1\textwidth]{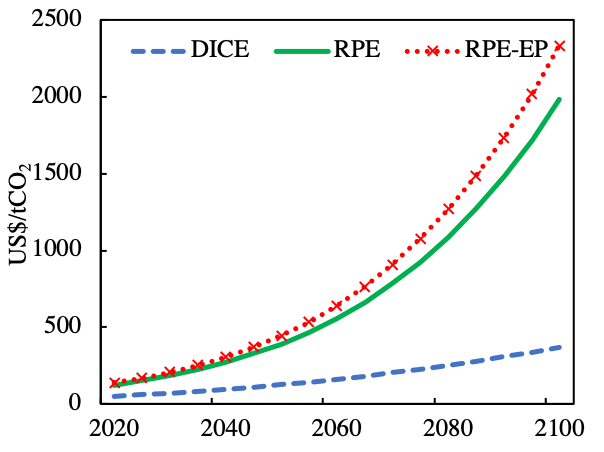}
        \centerline{\footnotesize{(c): Social cost of carbon}}
    \end{minipage}
    \begin{minipage}[t]{0.49\textwidth}
        \centering
        \includegraphics[width=1\textwidth]{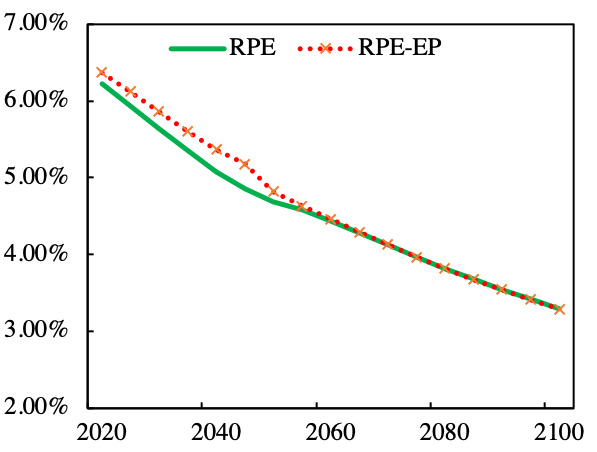}
        \centerline{\footnotesize{(d): Relative price effect}}
    \end{minipage}
    \begin{tablenotes}[flushleft]
    \footnotesize
    \item Notes: The dashed blue line represents the DICE model for comparison. The solid green line describes the model augmented with the relative price effect (RPE) as in \citet{drupp2021relative}. The dotted red line displays our model with both relative price effect and endogenous preference (RPE-EP).   
    \end{tablenotes}
\end{threeparttable}
    \caption{Climate policy, relative price effect, and endogenous preference}
    \label{Fig:EP}
\end{figure} 

Figure \ref{Fig:EP} displays the optimal climate policy in the presence of endogenous preferences. Under the optimal climate policy, the atmospheric temperature rises by 2.50$^\circ C$ by 2100, which is 0.06$^\circ C$ below that in \citet{drupp2021relative} and 0.80\celsius{} lower than in DICE. Evidently, the relative price effect is a major force in limiting warming, while endogenous preferences are relatively minor.

Net zero is hit in 2055 under both exogenous and endogenous preferences. However, industrial carbon emission is discernibly lower under endogenous preferences. In 2050, unabated emissions come to 1.29 GtCO\textsubscript{2} if abatement policies are adjusted according to endogenous preferences, compared to 5.86 GtCO\textsubscript{2} under the relative price effect. Both cases significantly bring forward the timing of full de-carbonization in DICE, by roughly half a century.

Panel (c) shows that higher abatement efforts are driven by the increased social cost of carbon. If the social planner is aware that abatement policy can increase the social preference for non-market goods, the social cost of carbon in 2020 is 139 US\$/tCO\textsubscript{2}, which is 15  US\$/tCO\textsubscript{2} higher than under the relative price effect alone, and more than double that of DICE. In 2050, the social cost of carbon is 445 US\$/tCO\textsubscript{2} under endogenous preference, which is 54 US\$/tCO\textsubscript{2} higher than under the relative price effect alone, and more than triple that of DICE. By the end of this century, the social cost of carbon reaches 2231  US\$/tCO\textsubscript{2}. This can be attributed to both the increasing relative scarcity of non-market goods and the strengthened social preferences for them.

Panel (d) compares the relative price effects between our model and \citet{drupp2021relative}. The relative price effect in our model is higher than \citet{drupp2021relative} in the first half of the presented period, while the two lines overlap in the second half. That is because the optimal abatement rate approaches one and stabilizes at near total abatement\textemdash recall that DICE does not allow for negative emissions. Thus, social preferences as governed by Equation (\ref{eq:preference}) are also stay constant. 

\begin{figure}[htp]
\centering
\begin{threeparttable}
    \begin{minipage}[t]{0.49\textwidth}
        \centering
        \includegraphics[width=1\textwidth]{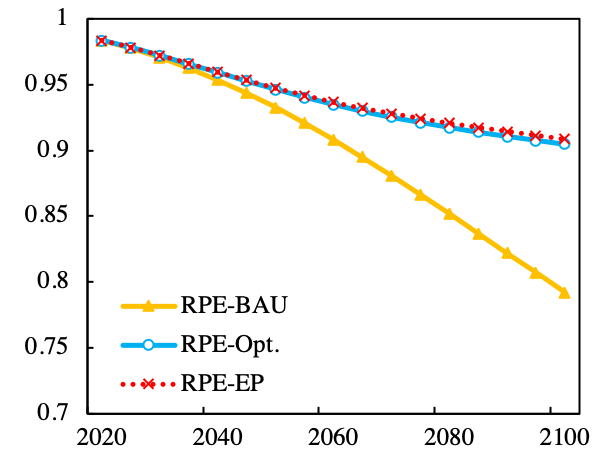}
        \centerline{\footnotesize{(a): Remaining stock of non-market goods}}
    \end{minipage}
    \begin{minipage}[t]{0.49\textwidth}
        \centering
        \includegraphics[width=1\textwidth]{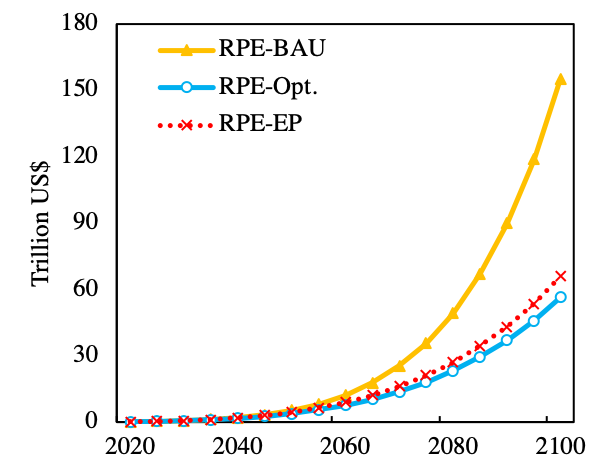}
        \centerline{\footnotesize{(b): Consumption-equivalent damages on NMG}}
    \end{minipage}
    \begin{tablenotes}[flushleft]
    \footnotesize
    \item Notes: NMG is short for non-market goods. RPE is short for relative price effect. BAU is short for business as usual scenario and Opt. stands for optimal scenario. RPE-EP denotes the relative price effect with endogenous preferences and adopts optimal climate policy. Panel (a) shows the remaining stock of non-market goods as a fraction of the original stock. Panel (b) displays the monetary value of the damage to non-market goods. 
    \end{tablenotes}
\end{threeparttable}
    \caption{Climate impact on non-market goods}
    \label{Fig:NMG}
\end{figure} 

Panel (a) of Figure \ref{Fig:NMG} shows that the volume of non-market goods will decrease by 9.12\% in 2100 after accounting for endogenous preference under optimal climate policy. By comparison, the volume of non-market goods suffers climate damage of 9.51\% under optimal abatement, and of 20.80\% in the business as usual (BAU) scenario. Because the damage level is driven by temperature rise that is barely different with and without endogenous preference, the destroyed stock of non-market goods is only slightly lower in the former situation.

We show the monetary value of the destroyed non-market goods in Panel (b). If optimal climate policy that accounts for endogenous preference is implemented, non-market goods damages amount to a consumption loss in market goods\footnote{To compute this, one should first obtain the absolute damage level of non-market goods, which is $E_0-E_t$. Then, this reduction in non-market goods stock should be multiplied by the relative price of non-market goods, as dictated by Eq.(\ref{eq:wtp}).} of 4.6 trillion US\$ in 2050, and of 65.9 trillion US\$ in 2100. By comparison, the market goods consumption derived from the model is 221.2 trillion US\$ and 571.7 trillion US\$, respectively. Thus, non-market goods damages are equivalent to a consumption loss in market goods by 2.1\% and 11.5\% in 2050 and 2100, respectively. 

Although non-market goods decline by less in volume under endogenous preferences, their monetary value is nonetheless higher. The reason lies in that the abatement policy increases the social preference for non-market goods and therefore increases its relative price further. With exogenous preferences, non-market damages are equivalent to a reduction in market consumption of 3.9 trillion US\$ in 2050 and 56.5 trillion US\$ in 2100, 0.7 trillion US\$ and 9.4 trillion US\$ lower, respectively, than with endogenous preferences. These results show that neglecting social preference changes leads to a non-trivial underestimate (by about 15\% in both 2050 and 2100) of non-market goods damages.

Most strikingly, under the business as usual scenario, damages on non-market goods will climb to an equivalent loss in market consumption of 155.0 trillion US\$ in 2100. Both higher levels of absolute climate damage (as shown in Panel (a)) and higher relative prices contribute to this result.

\subsection{Sensitivity}
We perform sensitivity analysis to show how $\beta_\mu$, the impact of abatement policy on preferences, influences optimal climate policy. If a value of zero is chosen, the model replicates the results of \citet{drupp2021relative}. For a value of 0.02, it generates the baseline result above.

\begin{figure}[htp]
\centering
\begin{threeparttable}
    \begin{minipage}[t]{0.49\textwidth}
        \centering
        \includegraphics[width=1\textwidth]{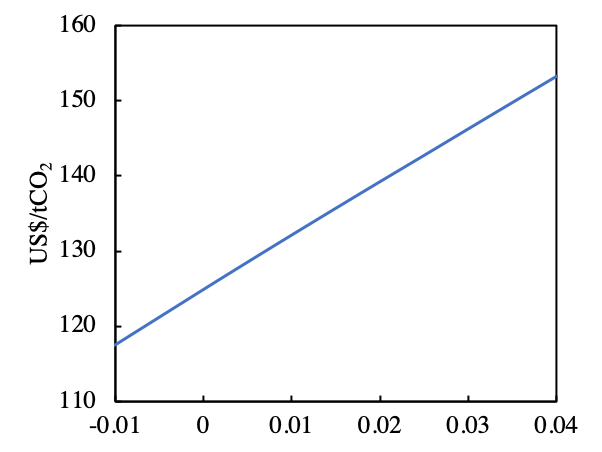}
        \centerline{\footnotesize{(a): Social cost of carbon 2020}}
    \end{minipage}
    \begin{minipage}[t]{0.49\textwidth}
        \centering
        \includegraphics[width=1\textwidth]{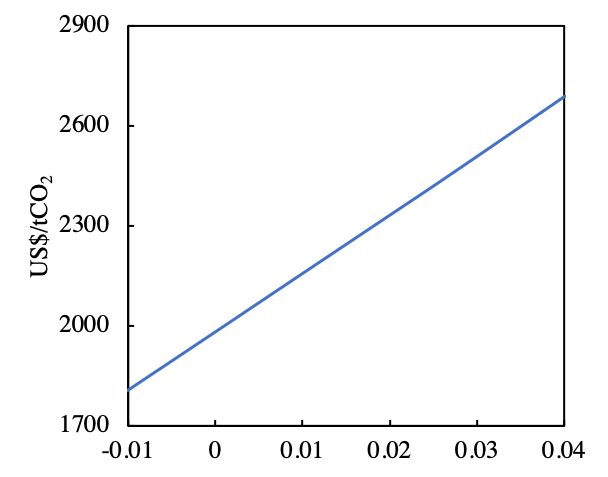}
        \centerline{\footnotesize{(b): Social cost of carbon 2100}}
    \end{minipage}
    \begin{minipage}[t]{0.49\textwidth}
        \centering
        \includegraphics[width=1\textwidth]{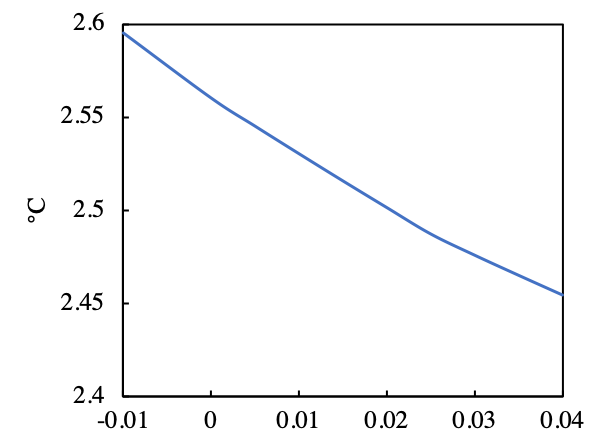}
        \centerline{\footnotesize{(c): Temperature rise 2100}}
    \end{minipage}
    \begin{minipage}[t]{0.49\textwidth}
        \centering
        \includegraphics[width=1\textwidth]{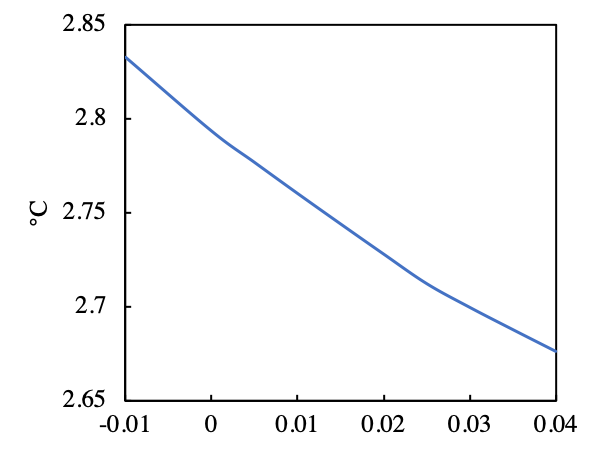}
        \centerline{\footnotesize{(d): Peak temperature rise}}
    \end{minipage}
    \begin{tablenotes}[flushleft]
    \footnotesize
    \item Notes: The horizontal axis represents the ability of policies to alter social preferences; $\beta_\mu=0.2$ is the benchmark.
    \end{tablenotes}
\end{threeparttable}
    \caption{Sensitivity tests on policy response}
    \label{Fig:Sensitivity}
\end{figure} 

As depicted in Figure \ref{Fig:Sensitivity}, the social cost of carbon is more sensitive to preference response parameter $\beta_\mu$ than is warming. When $\beta_\mu$ varies from 0 to 0.04, the social cost of carbon in 2020 increases by 23\% from 125 US\$/tCO\textsubscript{2} to 153 \$/tCO\textsubscript{2}, and in 2100 will increase by 36\% from 1981 US\$/tCO\textsubscript{2} to 2687 US\$/tCO\textsubscript{2}. Compared to fixed social preference, both temperature rise in 2100 and peak temperature rise decline by roughly 0.1 $^\circ C$ if the preference response parameter is 0.04.

For the extreme case $\beta_\mu=-0.01$, as improper policies make people resent preserving non-market goods, the social cost of carbon is reduced compared to without preference change, to 118 US\$/tCO\textsubscript{2} in 2020 and 1806 US\$ in 2100. However, its values in this case remain far above the levels revealed by DICE (51 US\$/tCO\textsubscript{2} and 370 US\$/tCO\textsubscript{2}). The scarcity effect dominates the preference effect.

\begin{table}[htbp]
  \centering
  \caption{Sensitivity of social cost of carbon (\$/tCO\textsubscript{2}) in 2100 }
\begin{threeparttable}
    \begin{tabular}{lcccccc}
    \toprule
    $\theta$ & $\beta_\mu=-0.01$  & $\beta_\mu=0$   &  $\beta_\mu=0.01$ &  $\beta_\mu=0.02$  & $\beta_\mu=0.03$ & $\beta_\mu=0.04$ \\
    \midrule
    0.29 & 22302 & 24759   & 27262  &  29804   & 32382   & 34999  \\
    0.5 & 1806 & 1981   & 2156  & 2331  & 2508  & 2687  \\
    1 & 328 & 345    & 363    & 381    & 400    & 420  \\
    1.3 & 272 & 281&  290& 300 & 310 & 320\\
    2 & 238& 242    & 246   & 251    & 255   & 260  \\
    $\infty$ &216& 217    & 218    & 219    & 221   & 222  \\
    \bottomrule
    \end{tabular}%
    \begin{tablenotes}[flushleft]
    \footnotesize
    \item Notes: $\theta$ is the substitution elasticity between two types of consumption in the utility function (\ref{eq:lifetime utility}). $\beta_\mu$ is the preference response parameter to the abatement rate. ADD $\beta_\mu = -0.01$ $\theta = 1.3$
    \end{tablenotes}
\end{threeparttable}
  \label{tab:sen}%
\end{table}%

Table \ref{tab:sen} displays the social cost of carbon under different combinations of parameter values. A substitution elasticity $\theta = 0.29$ is the lowest value considered in \citet{drupp2021relative}, who surveyed some empirical studies on this parameter. 
\citet{drupp2021relative} found a mean value of 1.30 for $\theta$, mild substitution. They acknowledged that this value is in contrast with that commonly-assumed complementarity, $\theta < 1$, in theoretical studies. 

If market and non-market goods are less substitutable, the preference response parameter is more capable of altering long-run climate policy. When the two are perfect substitutes, choosing the highest value for preference response only increases the social cost of carbon from fixed preferences by 2.4\%. For $\theta = 1$, the social cost of carbon in 2100 varies by roughly 10\% around the central choice of $\beta_\mu = 0.02$. If the lowest value $\theta = 0.29$ is chosen, variations are even more pronounced, up and down by nearly 17\% compared to the central choice of $\beta_\mu$. Our baseline calibration for $\theta = 0.5$ generates results lying in between.

\section{Conclusion and discussion\label{sec:conclusion}}

This paper examines how optimal carbon abatement decision changes if social preferences for non-market goods respond to abatement. We find that in addition to the scarcity effect, endogenous social preferences increases the relative price of non-market goods. The social cost of carbon increases correspondingly. Compared to fixed social preference, the social cost of carbon is elevated by 54 US\$/tCO\textsubscript{2} to 445 US\$/tCO\textsubscript{2} in 2050, and by 350 US\$/tCO\textsubscript{2} to 2331 US\$/tCO\textsubscript{2} in 2100. If abatement policies are properly introduced to reflect these costs, the climate impact on non-market goods amounts to a consumption loss in market goods of 3.9 and 56.5 trillion US\$ in 2050 and 2100, respectively. If no such policies are implemented, however, non-market goods damages will almost triple in monetary value. 

The paper focuses on the benefit end of climate governance. \citet{nyborg2016social} noted that social norms can be important in turning a vicious cycle into a virtuous cycle, or vice versa. In line with this spirit, \citet{konc2021social} and \citet{mattauch2022economics} showed how social preference for products with different carbon intensity contributes to abatement. Recently, \citet{besley2023political} further showed that changing social preference can be complementary to technological advances, fueling the net-zero transition. Although changes in social preferences are exogenous in their study, their results shed light on the previously-overlooked role of social preference favorable to deep carbon reduction. However, preference changes not only occur among goods with different pollution intensities, but also among those subject to different levels of climate damage. The former speaks to how much carbon we can abate, the latter governs how we value climate damages\textemdash that is the main focus of this study. Both of them matter to climate policy decisions.

There are three interesting avenues to explore along this paper. First, calibrating the preference response parameter entails empirical evidence that is currently lacking. To address this concern, our study performs sensitivity tests as a preliminary attempt. However, the impact of endogenous preference is still highly uncertain, and demands more efforts from both empirical literature as well as utilizing advanced probability tools.

Second, there are several other channels via which social preferences are endogenous in the complex climate-economy system. For example, in the case of market and non-market goods, household demand for non-market goods increases with affluence. In other words, social preference is responsive to the income level. Also, the appreciation of non-market goods can plausibly be affected by the climate system, if any catastrophic climate events raise public awareness of the value of non-market goods.

Last, we abstract from other values of non-market goods in the model. For example, non-market goods are considered as a necessary input in economic production, particularly in developing countries. All this is deferred to future research.

\bibliographystyle{elsarticle-harv} 
\bibliography{cas-refs}





\end{document}